\documentclass[showpacs,twocolumn,preprintnumbers,amsmath,amssymb]{revtex4}

\usepackage{graphicx}
\usepackage{dcolumn}
\usepackage{bm}

\begin{document}

\preprint{APS/123-QED}

\title{Characterizing and modeling cyclic behavior in non-stationary time series through
multi-resolution analysis}

\author{Dilip P. Ahalpara}
\email{dilip@ipr.res.in} \affiliation{Institute for Plasma Research, Near
Indira Bridge, Gandhinagar-382428, India}

\author{Amit Verma}
\email{amit.verma@yahoo.com}
\author{Prasanta K. Panigrahi}
\email{prasanta@prl.res.in}
\author{Jitendra C. Parikh}
\email{parikh@prl.res.in}

\affiliation{Physical Research Laboratory, Navrangpura, Ahmedabad-380009,
India}

\date{\today}

\begin{abstract}
A  method  based  on wavelet transform and genetic programming is proposed
for   characterizing   and  modeling  variations  at  multiple  scales  in
non-stationary  time  series. The cyclic variations, extracted by wavelets
and  smoothened by cubic splines, are well captured by genetic programming
in  the  form  of dynamical equations. For the purpose of illustration, we
analyze two different non-stationary financial time series, S\&P CNX Nifty
closing  index  of  the  National  Stock  Exchange  (India)  and Dow Jones
industrial   average   closing   values  through  Haar,  Daubechies-4  and
continuous  Morlet  wavelets for studying the character of fluctuations at
different  scales,  before modeling the cyclic behavior through GP. Cyclic
variations  emerge  at  intermediate  time  scales  and  the corresponding
dynamical  equations  reveal  characteristic behavior at different scales.
\end{abstract}

\maketitle

\section{\label{sec:Introduction}Introduction}
A number of non-stationary time series are known to comprise of
fluctuations having both stochastic and cyclic or periodic components.
Isolating fluctuations from these time series, at different scales for the
purpose of characterization and modeling is a research area of significant
interest \cite{Fract:Mandelbrot, Fract:Mantegna, Fract:Mankiw}. Here we
explicate a wavelet based approach for separating structured variations
from the stochastic ones in a time series before modeling them through
genetic programming (GP). For the purpose of illustration we have chosen
two financial time series, since the same is well-known to exhibit random
and structured behavior at different scales
\cite{Wavelet:Ram,Wavelet:Panigrahi}. At small time scales, the
fluctuations are primarily stochastic in nature; at higher scales the
random part is averaged out and characteristic nature of the variations
become transparent. For a reliable analysis one also needs to have a
reasonably long data set for which the financial time series are well
suited. The chosen time series are S\&P CNX Nifty closing index of the
National Stock Exchange (India) and Dow Jones industrial average closing
values, representing two sufficiently different economic climates so as to
bring out the efficacy of the present approach.

Study  and  characterization of fluctuations in financial time series have
been studied through a variety of approaches. For example, variations have
been  analyzed  through Levy-stable non-Gaussian model \cite{Levy:Schulz}.
Stochastic  nature  of  the  high  frequency  fluctuations and presence of
structured  behavior have emerged through study of these time series using
random  matrix  theory.  In particular, analysis of the cross-correlations
between         different         stocks        \cite{RandomMatrix:Plerou,
DelayRandomMatrix:Amritkar}  reveal universal and non-universal phenomena.
The  latter  ones indicate correlated behavior between stocks of different
companies.  This  behavior  can  manifest  in  the  composite  stock price
indices,  where the correlated behavior of several companies can give rise
to  structured  or  cyclic  behavior  in  appropriate time scales. Wavelet
transform,   because   of   its   multi-resolution  analysis  property  is
well-suited  to  isolate  fluctuations  and variations at different scales
\cite{Wavelet:Daubechies,Wavelet:Gencay}.

The goal of the present article is to demonstrate the usefulness of
combining wavelet transform with tools like genetic programming for
modeling of fluctuations. We carry out our analysis on two different
financial time series, in order to find out the similarities and
differences between them, from the perspective of fluctuations. Apart from
sharp transients representing sudden variations ascribable to physical
causes, the high frequency fluctuations at small scales are primarily
random in character. In all the time series, cyclic behavior emerges at
higher scales, when the random fluctuations are averaged out. The physical
nature of the cyclic phenomena is substantiated through both discrete and
continuous wavelets. In case of continuous wavelets, the scalogram clearly
reveals cyclic behavior at intermediate scales.  We then proceed to model
these variations through Genetic Programming (GP)
\cite{GA:Fog,GA:Holl,GA:Gold,GA:Melanie,GP:Szpiro}. For that purpose, we
smoothen the cyclic behavior at every scale, corresponding to different
levels of wavelet decomposition, through a cubic spline. It needs to be
mentioned that, since the purpose is to model cyclic behavior, physically
it is meaningful to smoothen the same, before trying to model them.

\begin{figure}
\centering
    {\resizebox{!}{5.5cm}{%
       \includegraphics{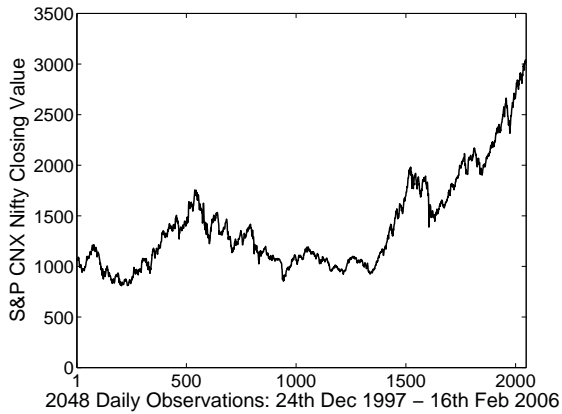}}}
\vspace{-0.1in} \caption{\label{fig:nifty2048_24dec1997}S\&P CNX Nifty
closing index data having 2048 points covering the daily index lying
within 24-Dec-1997 to 16-Feb-2006.}
\end{figure}

\begin{figure}
\centering
    {\resizebox{7.5cm}{10cm}{%
       \includegraphics{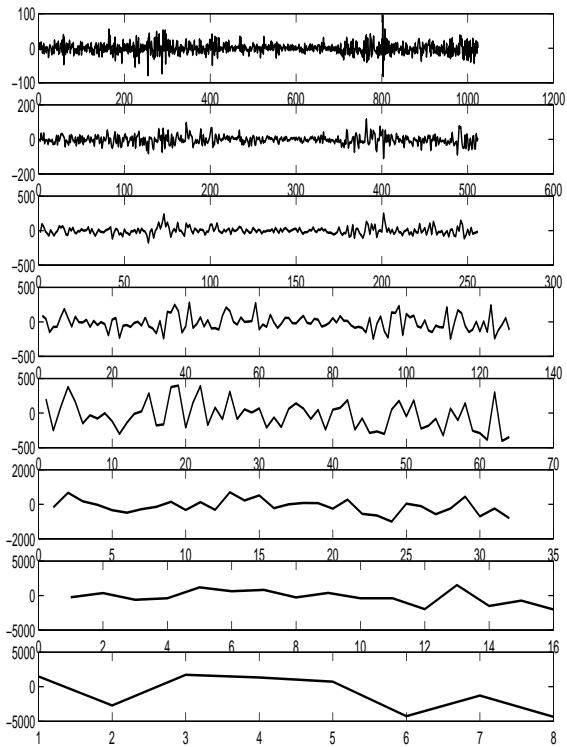}}}
\vspace{-0.1in} \caption{\label{fig:nifty2048_24dec1997_haar_all_hp}Haar
wavelet fluctuation coefficients for levels 1 to 8 for Nifty data.
Transient and stochastic behavior at small scales and ordered variations
at higher scales are evident.}
\end{figure}

\begin{figure}
\centering
    {\resizebox{!}{5.5cm}{%
       \includegraphics{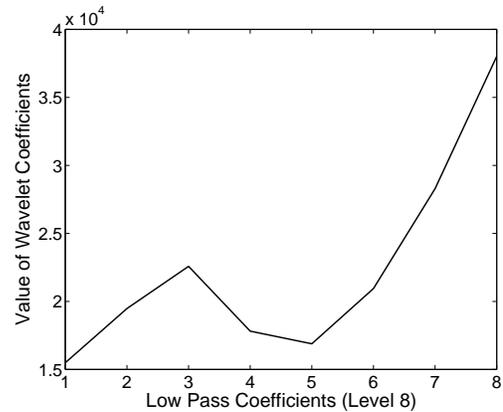}}}
\vspace{-0.1in} \caption{\label{fig:nifty2048_24dec1997_haar_lp}Haar
wavelet low pass coefficients for level 8 for Nifty data. As expected,
these coefficients resemble the average behavior of the time series.}
\end{figure}

\begin{figure}
\centering
    {\resizebox{!}{5.5cm}{%
       \includegraphics{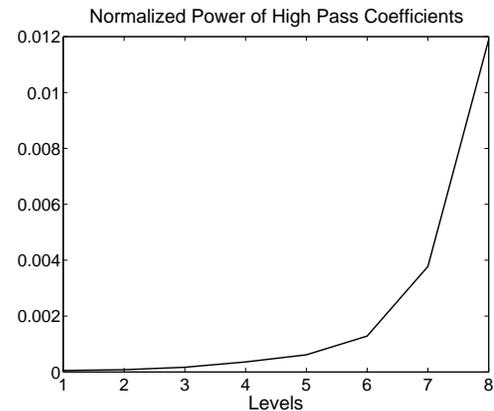}}}
\vspace{-0.1in}
\caption{\label{fig:nifty2048_24dec1997_haar_hp_pow}Normalized power of
Haar wavelet coefficients for different levels 1 to 8 for Nifty data.
Indication of rapid increase from $6^{th}$ level is clearly visible.}
\end{figure}

\begin{figure}
\centering
    {\resizebox{!}{5.5cm}{%
       \includegraphics{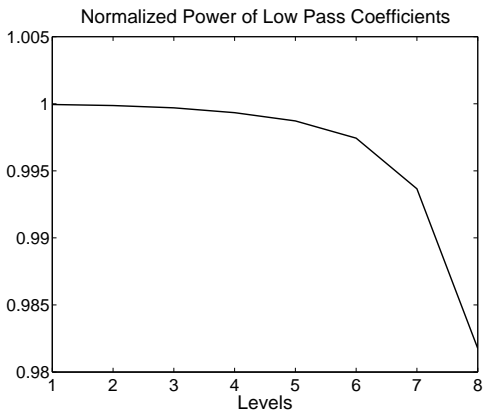}}}
\vspace{-0.1in}
\caption{\label{fig:nifty2048_24dec1997_haar_lp_pow}Normalized power of
Haar wavelet low pass coefficients for different levels 1 to 8 for Nifty
data. One observes decrease in low-pass power from $6^{th}$ level
onwards.}
\end{figure}

We study the fluctuation characteristics of two different financial time
series, S\&P CNX Nifty closing index of the National Stock Exchange
(India) and Dow Jones industrial average closing values, through wavelet
transforms belonging to both discrete and continuous families. Haar and
Daubechies-4 (Db4) from the discrete wavelet family and the continuous
Morlet wavelet are used to analyze the time series. As has been observed
earlier, at small scales, the fluctuations captured by the wavelet
coefficients exhibit self-similar character \cite{Fractal:Manimaran}.
Clear cyclic behavior emerges in medium scales and is evident from both
discrete and continuous wavelet analysis. It is found that, GP captures
the cyclic behavior at each scale quite well. The dynamical equations are
primarily linear with nonlinear additive terms of Pad\'{e} type. These
equations are checked for their predictive capabilities by making
out-of-sample predictions. One-step out-of-sample predictions are made
which use given time lagged values successively and predict the next data
set value. It is found that the one-step predictions are very good.

\begin{figure}
\centering
    {\resizebox{7.5cm}{10cm}{%
       \includegraphics{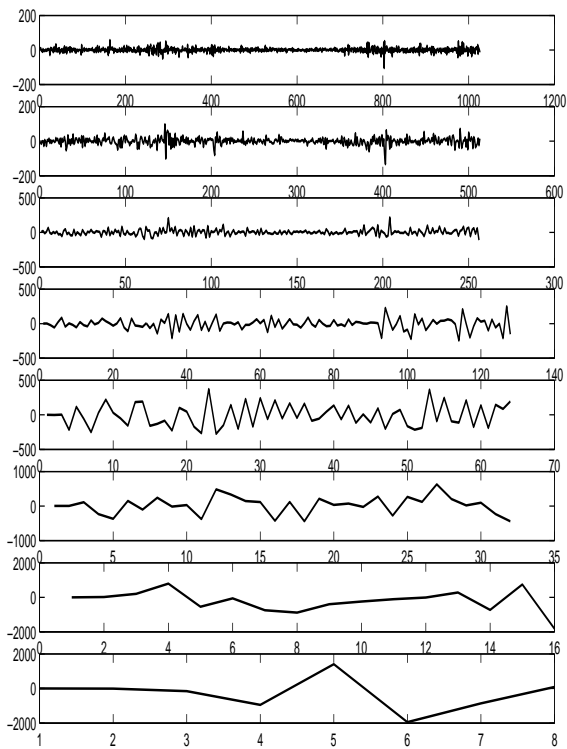}}}
\vspace{-0.1in} \caption{\label{fig:nifty2048_24dec1997_db4_all_hp}Db4
Wavelet coefficients for different levels 1 to 8 for Nifty data. Cyclic
behavior at intermediate scales are well captured by the wavelet
coefficients. This behavior is present both at local and global levels.}
\end{figure}

\begin{figure}
\centering
    {\resizebox{!}{5.5cm}{%
       \includegraphics{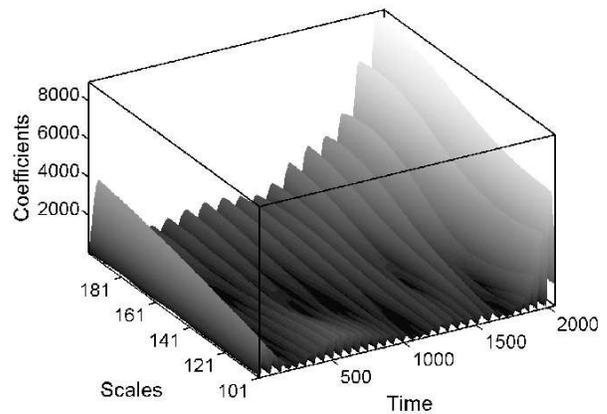}}}
\vspace{-0.1in}
\caption{\label{fig:nifty2048_24dec1997_cwt_morl_101_200.eps}Scalogram of
Morlet wavelet coefficients for scales 101-200 of S\&P CNX Nifty closing
index values.}
\end{figure}

\begin{figure}
\centering
    {\resizebox{!}{5.5cm}{%
       \includegraphics{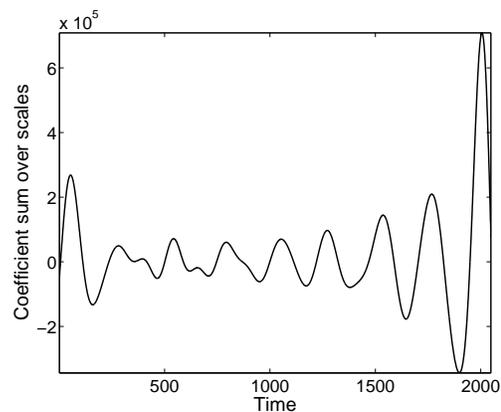}}}
\vspace{-0.1in}
\caption{\label{fig:nifty2048_24dec1997_cwt_morl_101_200_sum_scales.eps}Sum
of wavelet coefficients for Nifty data over all scales as a function of
time. An approximate periodic behavior with periodicity of about 200
trading days is evident.}
\end{figure}


The paper is organized as follows. In Section II, we give a brief
introduction to wavelets before carrying out wavelet decomposition of both
the data sets considered to study the character of the variations at
different scales. Cyclic variations at different scales is extracted
through Daubechies-4 wavelet and confirmed by continuous Morlet wavelet.
We then proceed to model the cyclic phenomenon through GP in section III
and conclude in section IV, after pointing out a number of applications
and future directions of work through the present method.

\section{Wavelet  Transform} Wavelet transform provides
a powerful tool for the analysis of transient and non-stationary data and
is particularly useful in picking out characteristic variations at
different resolutions or scales \cite{TimeSeries:Percival}. In the context
of financial time series \cite{Stoachistic:Clark,Forecasting:Poon}, it has
found extensive applications. It has been used for the study of commodity
prices \cite{Wavelet:Connor}, in measuring correlations
\cite{Wavelet:Simonsen}, in the study of foreign exchange rates
\cite{Wavelet:Karuppiah} and for predicting stock market behavior
\cite{Wavelet:Ramsey,Wavelet:Hayward}, to name a few. This linear
transform separates a data set in the form of low-pass or average
coefficients, resembling the data itself, and wavelet or high-pass
coefficients at different levels, which capture the variations at
corresponding scales. Wavelets can be continuous or discrete. In the
latter case, the basis elements are strictly finite in size, enabling them
to achieve localization, while disentangling characteristic variations at
different frequencies.

In discrete wavelet transform (DWT), the construction of basis set starts
with the scaling function $\varphi(x)$ (father wavelet) and the mother
wavelet $\psi(x)$, whose height and width are arbitrary:
$ \int \varphi dx=A ,~\int\psi dx=0,~ \int \varphi\psi dx=0, \int
|\varphi|^2 dx=1,~\int |\psi|^2dx=1$,  where $A$ is an arbitrary constant.
The scaling and wavelet functions, and their scaled translates, known as
daughter wavelets,
$ \psi_{j,k}=2^{j/2}\psi(2^j x-k)$,
are square integrable at different scales. Here, k and j respectively are
the translation and scaling parameters, with $-\infty\leq k \leq +\infty$.
Although conventionally, one starts with the scale value $j =0$, one can
begin from any finite value $j'$ and increase it by integral units. The
original mother wavelet corresponds to $\psi_{0,0}$. The daughter wavelets
are of a similar form as the mother wavelet, except that their width and
height differ by a factor of $2^{j}$ and $2^{j/2}$ respectively, at
successive levels. The translation unit $k/2^j$ is commensurate with the
thinner size of the daughter wavelet at scale $j$. In the limit
$j\rightarrow \infty$, these basis functions form a complete orthonormal
set, allowing us to expand a signal $f(t)$ in the form,
\begin{eqnarray} f(t)=\sum_{k=-\infty}^{+\infty}c_{j,k}
\varphi_{j,k}(t)+\sum_{k=-\infty}^{+\infty}\sum_{j'\geq j}^{}d_{j',k}
\psi_{j',k}(t)\end{eqnarray}
Here, $c_{j,k}$'s are the low-pass coefficients and $d_{j,k}$'s are the
high-pass or wavelet coefficients. They respectively capture the average
part and variations of the signal at scale $j$ and location $k$. For the
discrete wavelets,  the property of multi-resolution analysis (MRA) leads
to  $c_{j,k}=\sum_{n} h(n-2k)c_{j+1,n}, d_{j,k}=\sum_{n}
\tilde{h}(n-2k)c_{j+1,n},$ where $h(n)$ and ${\tilde{h}}(n)$ are
respectively the low-pass (scaling function) and high-pass (wavelet)
filter coefficients, which differ for different wavelets. Both low-pass
and high-pass coefficients at a scale $j$ can be obtained from the
low-pass coefficients at a higher scale ($c_{j+1,n}$). This implies that,
starting from the finest resolution of the signal,  one can construct both
scaling and wavelet coefficients, by convolution with the filter
coefficients $h(n)$ and ${\tilde{h}}(n)$.

For the Haar wavelet: $h(0)=h(1)=\frac{1}{\sqrt{2}}$ and
$\tilde{h}(0)=-\tilde{h}(1)=\frac{1}{\sqrt{2}}$. Haar basis is unique,
since it is the only wavelet, which is symmetric and compactly supported.
In a level one Haar wavelet decomposition, the level-I low-pass (average)
and high-pass (wavelet or detail) coefficients are respectively given by
the nearest neighbor averages and differences, with the normalization
factor of $\frac{1}{\sqrt{2}}$. In the subsequent step, the average
coefficients are divided into two parts, containing level-II high-pass and
level-II low-pass coefficients. The high-pass coefficients now represent
differences of averaged data points corresponding to a window size of two.
Wavelets belonging to Daubechies family are designed such that, the
wavelet coefficients are independent of polynomial trends in the data. For
example, Daubechies-4 wavelet satisfies, $ \int t \psi(t) dt=0$, in
addition to all other conditions. Because of this the wavelet coefficients
here capture fluctuations over and above the linear variations. For
Daubechies-4, $h(0)=-\tilde{h}(3)=\frac{1+\sqrt 3}{4\sqrt 2}$,
$h(1)=\tilde{h}(2)=\frac{3+\sqrt 3}{4\sqrt 2}$,
$h(2)=-\tilde{h}(1)=\frac{3-\sqrt 3}{4\sqrt 2}$,
$h(3)=-\tilde{h}(0)=\frac{1-\sqrt 3}{4\sqrt 2}$. We have used both Haar
and Daubechies-4 wavelets for isolating these fluctuations at different
scales and study their character. For continuous wavelet transform (CWT),
we have utilized Morlet wavelet, whose analyzing function is given by,
\begin{equation}
\psi(t)= {\pi}^{-1/4}e^{(-i\omega_0t-t^2/2)}.
\end{equation}
The corresponding wavelet coefficients are displayed as a function of
scale and time in a scalogram.

In DWT, a maximum of $N$ level decompositions can be carried out, when the
data set is of   size   $2^N$.   One   may   choose   to   have   a  less
number  of decompositions than $N$. Often one needs to supplement the data
with additional points to carry out  a  $N$  level  decomposition.  Both
in  DWT  and CWT, one encounters boundary  artifacts,  due  to  circular
or  other  forms  of  extensions. In  our  case,  for  minimizing  these
boundary  artifacts,  we have used symmetric    extension,   while
studying   the   behavior   of   wavelet coefficients.   The  variations
at  different  scales  are  characterized by   their   respective powers,
defined  as  the  squared  sum  of  the wavelet   coefficients at   that
level.  Since  in  wavelet  transform power  is  conserved, the  squared
sum  of  all  the  low  pass and high pass  coefficients add  up  to  the
squared  sum  of  the data points of the  time  series, called  as  the
total  power  or energy. We have used normalized  power which  is  the
power  at  a  given  level  divided  by the   total power.  Periodic
extension  has  been  used  for  analyzing the distribution  of  power  at
various  levels,  since  this  extension conserves  power.  The  power
plots  depicting  high  pass  power and low pass  power  clearly  reveal
the  character  of  the  fluctuation and the average behavior.

We start with the National Stock Exchange (NSE) of India Nifty daily
closing index values, consisting of 2048 data points, covering the
duration from $24^{th}$ December 1997 to $16^{th}$ February 2006. This
daily index is shown in Fig. \ref{fig:nifty2048_24dec1997}. We carry out
an eight level decomposition through Haar transform, since the same yields
a transparent picture about the nature of the variations. The high pass
coefficients are depicted in Fig.
\ref{fig:nifty2048_24dec1997_haar_all_hp}. Transient and cyclic behavior
at different scales are clearly visible. The corresponding low pass
coefficients corresponding to $8^{th}$ level are depicted in Fig.
\ref{fig:nifty2048_24dec1997_haar_lp}. As expected, these low pass
coefficients capture the average behavior of the time series (see Fig.
{\ref{fig:nifty2048_24dec1997}). At finer resolutions corresponding to
lower level wavelet coefficients, one can clearly see primarily random
nature of the fluctuations.

The non-random variations significantly increase from $6^{th}$ level
onwards. Substantial increase in power at $6^{th}$ level is also evident
in Fig. \ref{fig:nifty2048_24dec1997_haar_hp_pow}, which shows the
behavior of the normalized power of the high pass coefficients. The
normalized power of low pass coefficients shown in Fig.
\ref{fig:nifty2048_24dec1997_haar_lp_pow} start decreasing from $6^{th}$
level onwards, since the total power is conserved in wavelet transform. It
is worth mentioning that the $6^{th}$ level high pass coefficients
correspond to the differences of data points, averaged over a temporal
window size 32. A few transient phenomena are also revealed by these
coefficients. After sufficient averaging, cyclic behavior is seen to
emerge. The averaged low pass coefficients reveal a linear trend like the
time series in Fig. \ref{fig:nifty2048_24dec1997}; this can affect the
high pass coefficients. In order to remove this and capture the
characteristic nature of the variations, we have carried out decomposition
through Daubechies-4 wavelets. The structured and cyclic behavior is
transparently demonstrated in Fig.
\ref{fig:nifty2048_24dec1997_db4_all_hp}. This justifies the use of
Daubechies-4 wavelets, which removes linear trend from the high-pass
coefficients. Plots of both high-pass and low-pass normalized power reveal
that at higher scales fluctuations capture most of the energy; the
high-pass power increases, while that of low-pass decreases progressively.
CWT through Morlet wavelets also reveal the cyclic behavior at
intermediate scales as seen in Fig.
\ref{fig:nifty2048_24dec1997_cwt_morl_101_200.eps}. The scale values
indicate the window size of the Morlet wavelet corresponding to the same
number of days. We depict in Fig.
\ref{fig:nifty2048_24dec1997_cwt_morl_101_200_sum_scales.eps} sum over the
continuous wavelet coefficients over scales as a function of time. An
approximate periodic behavior is seen with a period of about 200 trading
days. The fact that wavelet coefficients containing both positive and
negative values add up to yield a periodic behavior indicates the presence
of correlated behavior. As is clear, purely random un-correlated
coefficients will not lead to this structure.

\begin{figure}
\centering
    {\resizebox{!}{5.5cm}{%
       \includegraphics{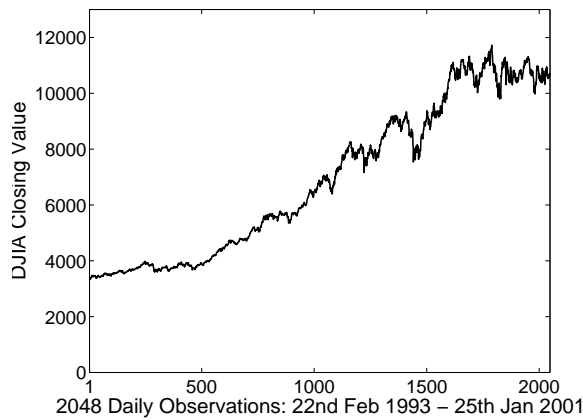}}}
\vspace{-0.1in} \caption{\label{fig:djia2048_22feb1993}DJIA closing values
having 2048 points covering the daily index lying within 22-Feb-1993 to
25-Jan-2001.}
\end{figure}

\begin{figure}
\centering
    {\resizebox{7.5cm}{10cm}{%
       \includegraphics{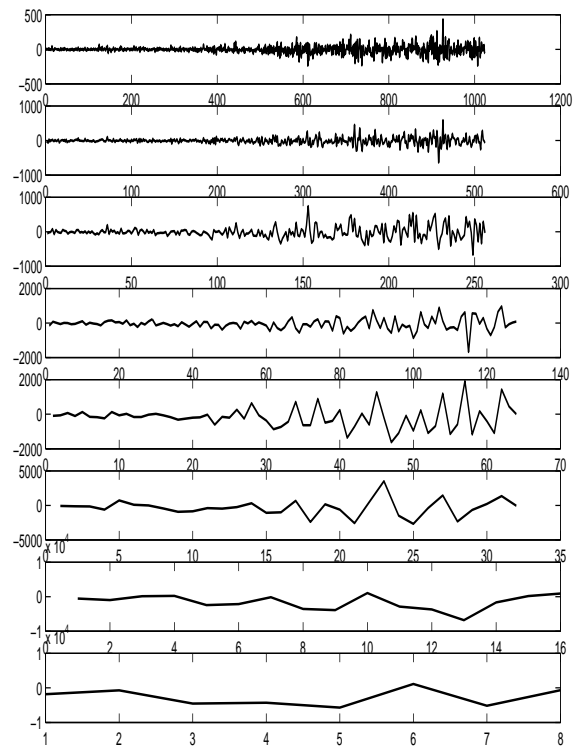}}}
\vspace{-0.1in} \caption{\label{fig:djia2048_22feb1993_haar_all_hp}Haar
wavelet coefficients for levels 1 to 8 for Dow Jones data.}
\end{figure}

\begin{figure}
\centering
    {\resizebox{!}{5.5cm}{%
       \includegraphics{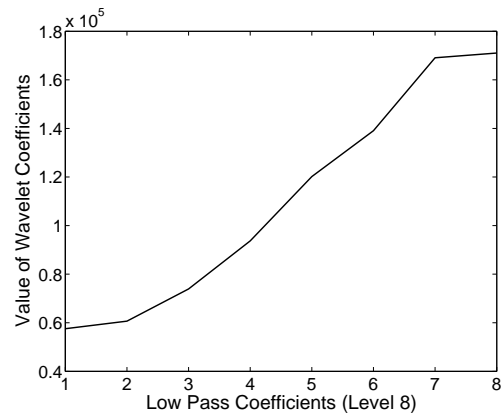}}}
\vspace{-0.1in} \caption{\label{fig:djia2048_22feb1993_haar_lp}Haar
wavelet low pass coefficients for levels 1 to 8 for Dow Jones data. A
linear trend is seen.}
\end{figure}

\begin{figure}
\centering
    {\resizebox{!}{5.5cm}{%
       \includegraphics{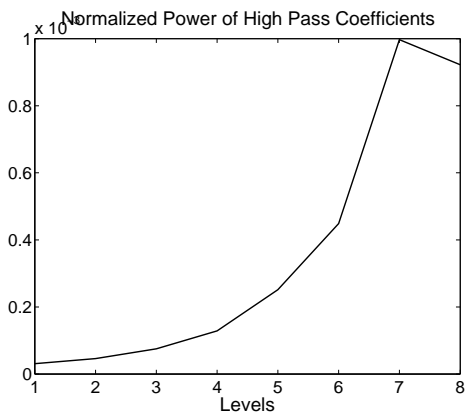}}}
\vspace{-0.1in}
\caption{\label{fig:djia2048_22feb1993_haar_hp_pow}Normalized power of
Haar wavelet high pass coefficients for levels 1 to 8 for Dow Jones data.
Significant increase in power is seen around $6^{th}$ level.}
\end{figure}

\begin{figure}
\centering
    {\resizebox{!}{5.5cm}{%
       \includegraphics{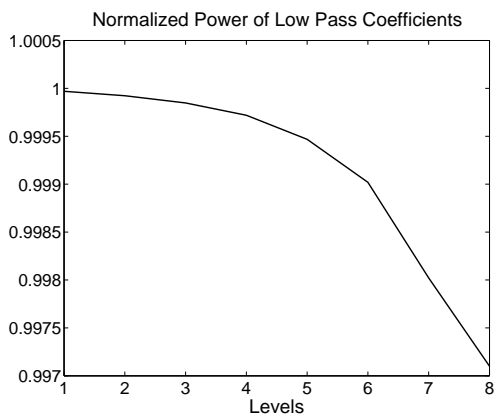}}}
\vspace{-0.1in}
\caption{\label{fig:djia2048_22feb1993_haar_lp_pow}Normalized power of
Haar wavelet low pass coefficients for levels 1 to 8 for Dow Jones data.
Complimenting the high-pass behavior, the low-pass power decreases rapidly
from $6^{th}$ level onwards.}
\end{figure}

\begin{figure}
\centering
    {\resizebox{7.5cm}{10cm}{%
       \includegraphics{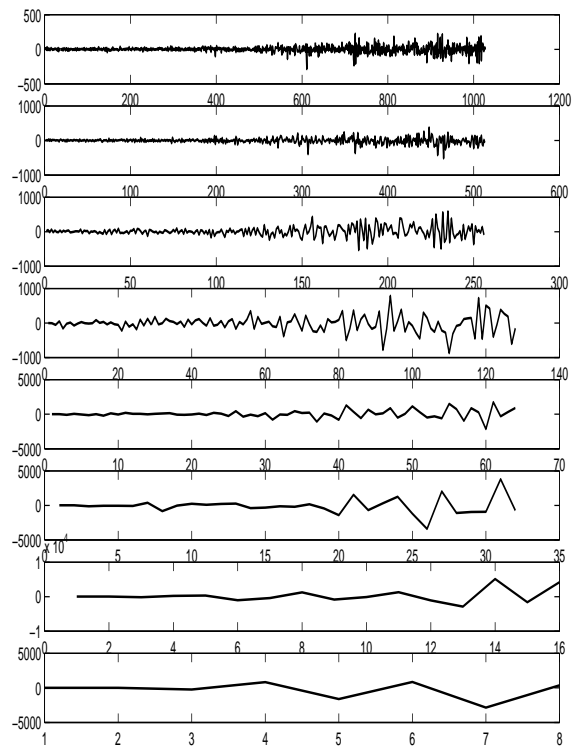}}}
\vspace{-0.1in} \caption{\label{fig:djia2048_22feb1993_db4_all_hp}Db4
wavelet coefficients for levels 1 to 8 for Dow Jones data. Significant
activity is seen in the second half of the wavelet coefficients.}
\end{figure}

We next consider the Dow Jones industrial average (DJIA) closing values,
shown in Fig. \ref{fig:djia2048_22feb1993} having 2048 data points for the
duration lying within $22^{nd}$ February 1993 to $25^{th}$ January 2001.
An eight level decomposition through Haar transform is then carried out to
infer about the nature of the variations. The high pass coefficients are
depicted in Fig. \ref{fig:djia2048_22feb1993_haar_all_hp}. Transient and
cyclic behavior at different scales are clearly visible like the previous
case of Nifty data. It is to be noted that the $2^{nd}$ half of the
forward wavelet coefficients for each level is having higher amplitudes as
compared to the first half of the coefficients. In this respect,
variations of wavelet coefficients of Nifty data and DJIA data have
different characteristics. The corresponding low pass coefficients are
shown in Fig. \ref{fig:djia2048_22feb1993_haar_lp}. As is evident, these
low pass coefficients show the trend of the time series. The normalized
power of the high pass coefficients (Fig.
\ref{fig:djia2048_22feb1993_haar_hp_pow}) shows a continuous rise with
levels having a small decline at level 8. Correspondingly the low pass
coefficients show continuous decrease in the power (Fig.
\ref{fig:djia2048_22feb1993_haar_lp_pow}).   In Fig.
\ref{fig:djia2048_22feb1993_db4_all_hp} we show the Db4 wavelet
coefficients for levels 1 to 8 for the purpose of comparison with the
corresponding Nifty index behavior.

\begin{figure}
\centering
    {\resizebox{!}{5.5cm}{%
       \includegraphics{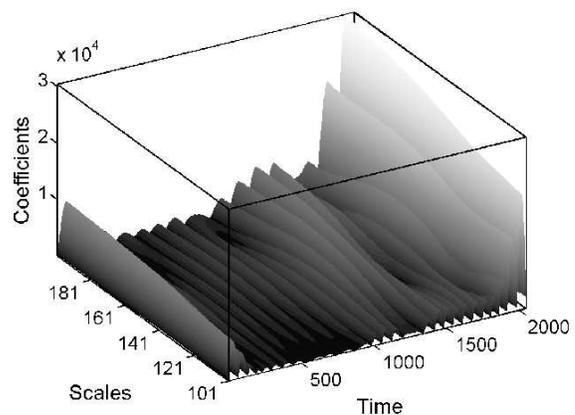}}}
\vspace{-0.1in}
\caption{\label{fig:djia2048_22feb1993_cwt_morl_101_200.eps}Scalogram of
Morlet Wavelet coefficients for scales 101-200 of DJIA closing values
indicating a cyclic behavior.}
\end{figure}

\begin{figure}
\centering
    {\resizebox{!}{5.5cm}{%
       \includegraphics{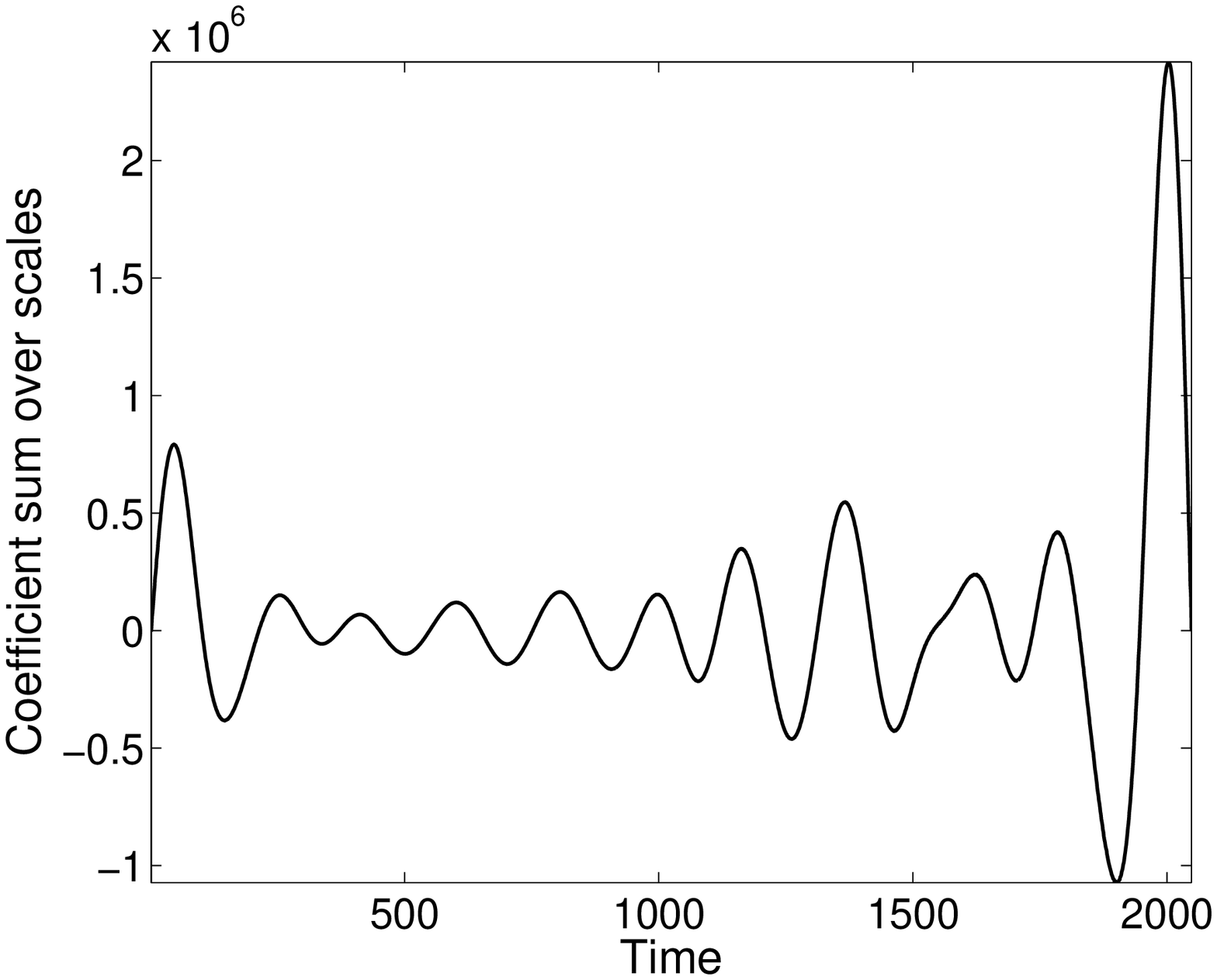}}}
\vspace{-0.1in}
\caption{\label{fig:djia2048_22feb1993_cwt_morl_101_200_sum_scales.eps}Sum
of wavelet coefficients for DJIA over all scales as a function of time. An
approximate cyclic behavior with periodicity of a duration of little less
than 200 trading days is evident.}
\end{figure}

We have studied the behavior of DJIA under CWT. The corresponding
scalogram is shown in Fig.
\ref{fig:djia2048_22feb1993_cwt_morl_101_200.eps}. Akin to the Nifty case
one sees cyclic behavior in the scale range 100 to 200. The sum of the
wavelet coefficients at all scales plotted as a function of time (Fig.
\ref{fig:djia2048_22feb1993_cwt_morl_101_200_sum_scales.eps}) reveal a
periodic behavior of little less than 200 trading days. In this context
both the financial time series show similar behavior. However the DJIA
time series is showing a tendency of bursty behavior which is absent in
Nifty case.

Considering the cyclic behavior of variations at intermediate scales, it
is interesting to see how well these wavelet coefficients can be analyzed
through the techniques of Genetic Programming in which the model equations
are built in the reconstructed phase space. This modeling can reveal
characteristic behavior of fluctuations at different scales.

\section{Modeling Cyclic Wavelet Coefficients through Genetic Programming}
In Genetic Programming one assumes the map equation relating time lagged
variables with the entity $X_{t+1}$ to be of the form
\cite{Fract:Grassb,GP:Szpiro},
\begin{equation}
X_{t+1} = f(X_{t}, X_{t-\tau}, X_{t-2\tau}, ... X_{t-(d-1)\tau})
\label{eq:MapEquation}
\end{equation}
Here f represents a function involving time series values $X_{t}$ in the
immediate past and arithmetic operators (+, -, $\times$ and $\div$). The
numbers appearing in function f are bounded between the range [-N, N],
where N is an integer number, we have chosen N to be 10. The numbers
within the above range are generated with the precision of 1 digit after
decimal point. In the above equation, d represents the number of previous
time series values that may appear in the function and $\tau$ represents a
time delay.

During the GP optimization, one considers a pool of chromosomes. A
chromosome represents a potential solution. Evolution to successive
generations is then carried out stochastically by applying genetic
operators, namely copy, crossover and mutation.

The sum of squared errors,
\begin{equation}
\bigtriangleup^{2} = \sum_{i=1}^{i=N} (X_{i}^{calc} - X_{i}^{given})^{2},
\label{eq:SumSqErrors}
\end{equation}
is minimized, where N represents number of $X_{t}$ values (Eq.
\ref{eq:MapEquation}) that are fitted during the GP optimization.

For a given chromosome, the lower the above sum of squared errors, the
better is the fit generated and therefore the corresponding chromosome
fairs better chance of participating in further evolutionary process
through its fitness measure. The fitness measure is derived from
$\bigtriangleup^{2}$ and is defined as in Eq. \ref{eq:RSquare}:
\begin{equation}
R^{2} = 1 - \frac{\bigtriangleup^{2}}{\displaystyle\sum_{i=1}^{i=N}
(X_{i}^{given} - \overline{X_{i}^{given}})^{2}}, \label{eq:RSquare}
\end{equation}
where $\overline{X_{i}^{given}}$ is the average of all $X_{i}$ (Eq.
\ref{eq:MapEquation}) to be fitted.

It is observed that during the optimization process, in order to get
higher and higher fitness measures, GP may lead to quite involved
chromosome strings. In order to discourage GP to over fit by generating
longer strings of chromosomes, the fitness measure is modified
\cite{GP:Szpiro} as follows,
\begin{equation}
r = 1 - (1 - R^{2})\frac{N-1}{N-k}, \label{eq:RSquareModified}
\end{equation}
where N is the number of equations to be fitted in the training set and k
is the total number of time lagged variables of the form $X_{t}$,
$X_{t-\tau}$, $X_{t-2\tau}$, ... etc (including repetitions) occurring in
the given chromosome. This modified fitness measure prefers a parsimonious
model by generating crisp map equations for chromosomes. For $R^{2}$ close
to 0, the modified fitness measure r can be negative.

The map equations generated by above GP prescription is then used to make
out-of-sample predictions outside the fitted set of data and the measure
of goodness of predictions is ascertained by normalized mean square
error(NMSE) as given by Eq. \ref{eqn:NMSE},
\begin{eqnarray}
NMSE = \frac{1}{N} \frac{\sum_{i=1}^{i=N} (X_{i}^{calc} -
X_{i}^{given})^{2}}{variance ~of ~N ~data ~points} \label{eqn:NMSE}
\end{eqnarray}

On trying to model these wavelet coefficients, it is found that due to
sharp variations, the GP optimization does not lead to convergence having
good fitness values. We have therefore found it necessary to smoothen
these wavelet coefficients using an appropriate method. For all the
wavelet coefficients of different levels considered, we smoothen them
using a cubic Spline interpolation method. We generate smoothened
coefficients by incorporating 4 additional points which are sampled by
cubic Spline method for each consecutive pair of points. The piecewise
polynomial form of the smoothened data is well suited with a similar
structure used for the map equation (Eq. \ref{eq:MapEquation}) searched by
Genetic Programming. It is worth emphasizing that this procedure is
appropriate since we are modeling cyclic behavior which are generally
smooth in nature as compared to sharp variations of transients.

We now proceed with modeling of smoothened Db4 forward wavelet
coefficients using GP at level=6, 7 and 8.

\subsection{Modeling variations in S\&P CNX Nifty closing index}
As seen earlier cyclic and structured variations emerge at relatively
higher scales. For modeling purpose, we consider the 6, 7 and $8^{th}$
level coefficients generated using Db4 forward wavelet transform. These
coefficients show considerable cyclic fluctuations both at local and
global scales. At $6^{th}$ level, bursty behavior is also seen.

\begin{figure}
\centering
    {\resizebox{!}{5.5cm}{%
       \includegraphics{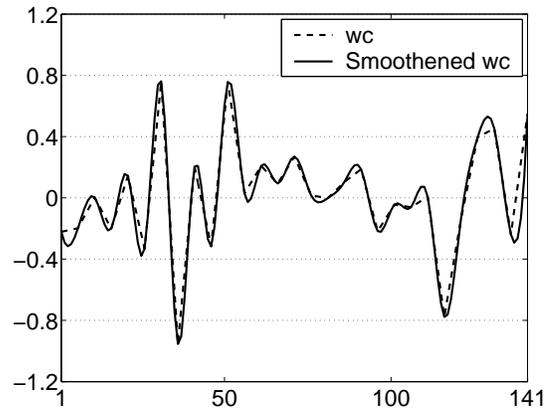}}}
\vspace{-0.1in} \caption{\label{fig:Nifty_cD6_CompareSpline141}Db4 wavelet
coefficients (wc) and Spline interpolated wc for $6^{th}$ level for Nifty
data.}
\end{figure}

\begin{figure}
\centering
    {\resizebox{!}{5.5cm}{%
       \includegraphics{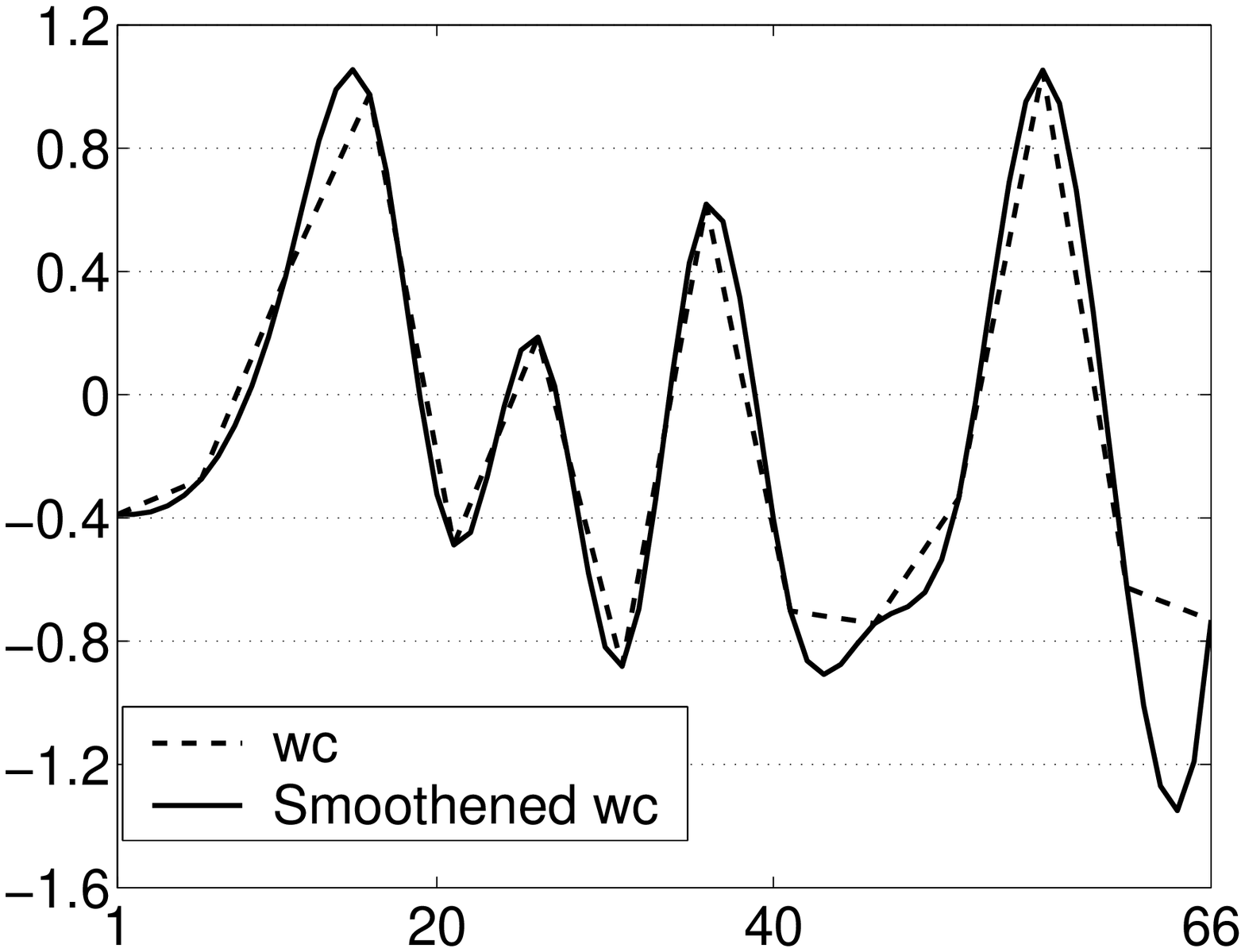}}}
\vspace{-0.1in} \caption{\label{fig:Nifty_cD7_CompareSpline66}Db4 wavelet
coefficients (wc) and Spline interpolated wc for $7^{th}$ level for Nifty
data.}
\end{figure}

\begin{figure}
\centering
    {\resizebox{!}{5.5cm}{%
       \includegraphics{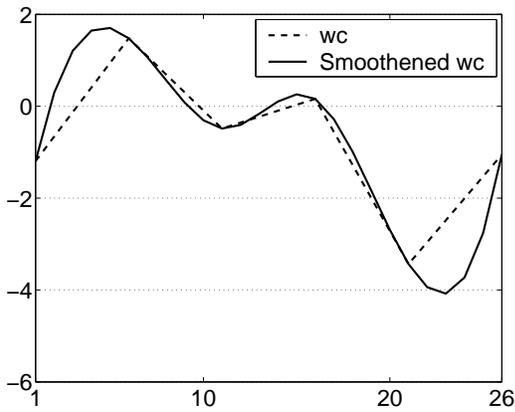}}}
\vspace{-0.1in} \caption{\label{fig:Nifty_cD8_CompareSpline26}Db4 wavelet
coefficients (wc) and Spline interpolated wc for $8^{th}$ level for Nifty
data.}
\end{figure}

We have divided the values of these wavelet fluctuations by 1000 for the
sake of computational convenience. The smoothened wavelet coefficients for
level=6, 7 and 8 are shown in Figs. \ref{fig:Nifty_cD6_CompareSpline141},
\ref{fig:Nifty_cD7_CompareSpline66} and
\ref{fig:Nifty_cD8_CompareSpline26} along with original wavelet
coefficients.

The smoothened wavelet coefficients are then modelled using GP. We have
used d=5 and $\tau$=$1$ for these fits and the resulting fits are very
good having fitness values 0.99499 (level=6), 0.99498 (level=7) and
0.997038 (level=8).

The map equations are shown in Eq. \ref{eqn:Nifty_6thLevel},
\ref{eqn:Nifty_7thLevel} and \ref{eqn:Nifty_8thLevel}.


\begin{tiny}
\begin{eqnarray}
X_{t+1}^{(Level=6)}&=&1.9762X_{t}+\frac{2.5298X_{t}}{X_{t-2\tau}+10.0}-1.5627X_{t-\tau}+0.3008X_{t-3\tau}\nonumber \\
&+&\frac{0.0006023X_{t-2\tau}}{X_{t-4\tau}-00339}
\label{eqn:Nifty_6thLevel}
\end{eqnarray}
\end{tiny}

\begin{tiny}
\begin{eqnarray}
X_{t+1}^{(Level=7)}=2.814X_{t}-2.9401X_{t-\tau}+1.1239X_{t-2\tau}-0.01761X_{t-3\tau}
\label{eqn:Nifty_7thLevel}
\end{eqnarray}
\end{tiny}

\begin{tiny}
\begin{eqnarray}
X_{t+1}^{(Level=8)}=1.5X_{t}-0.6522X_{t-2\tau}-\frac{0.368(X_{t-\tau}-2.9X_{t-4\tau}+1.3103)}{2.4X_{t}+3.18}
\label{eqn:Nifty_8thLevel}
\end{eqnarray}
\end{tiny}

The GP fit obtained by these equations are quite good and these are shown
for level=6, 7 and 8 in Figs. \ref{fig:Nifty_cD6_Fit130},
\ref{fig:Nifty_cD7_Fit55} and \ref{fig:Nifty_cD8_Fit18} respectively.

\begin{figure}
\centering
    {\resizebox{!}{5.5cm}{%
       \includegraphics{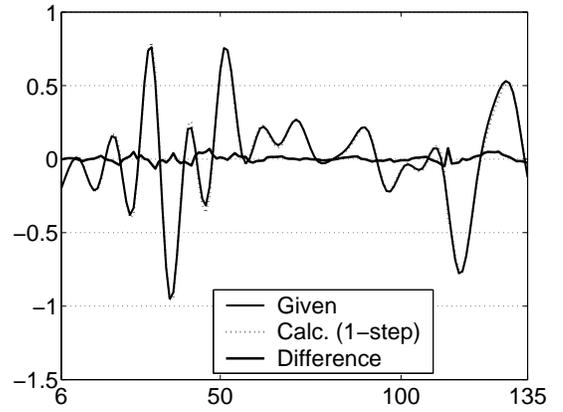}}}
\vspace{-0.1in} \caption{\label{fig:Nifty_cD6_Fit130}Fit for 130 data
points using GP solution for Db4 level-6 wavelet coefficients for Nifty
data.}
\end{figure}

\begin{figure}
\centering
    {\resizebox{!}{5.5cm}{%
       \includegraphics{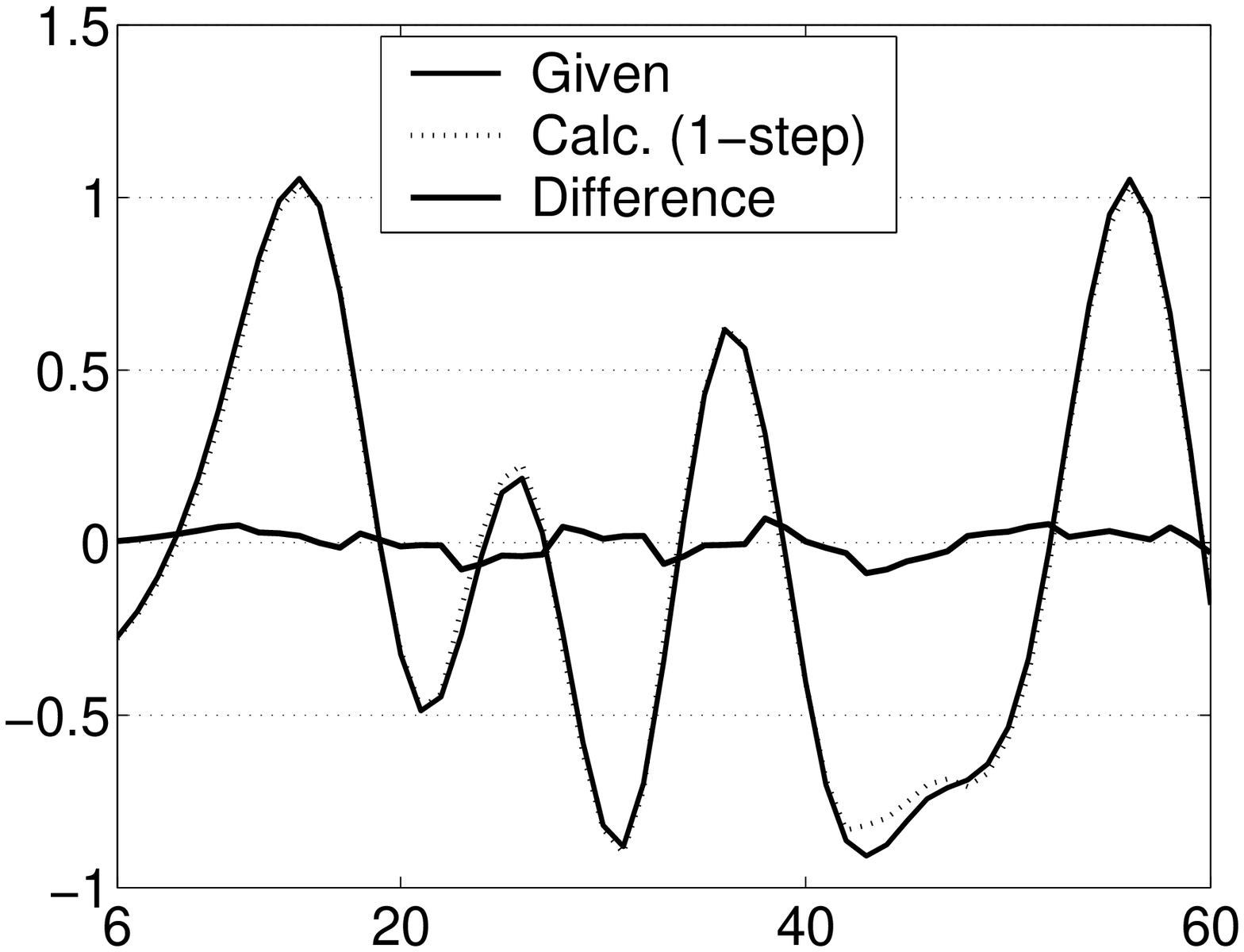}}}
\vspace{-0.1in} \caption{\label{fig:Nifty_cD7_Fit55}Fit for 55 data points
using GP solution for Db4 level-7 wavelet coefficients for Nifty data.}
\end{figure}

\begin{figure}
\centering
    {\resizebox{!}{5.5cm}{%
       \includegraphics{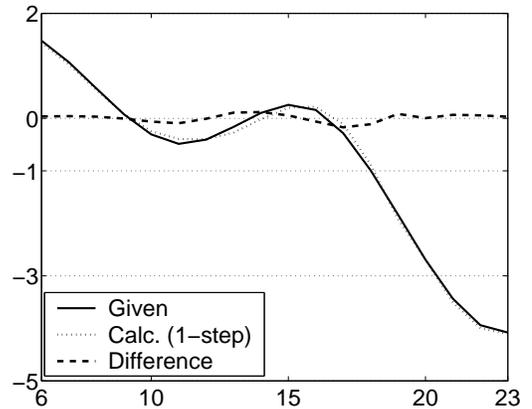}}}
\vspace{-0.1in} \caption{\label{fig:Nifty_cD8_Fit18}Fit for 18 data points
using GP solution for Db4 level-8 wavelet coefficients for Nifty data.}
\end{figure}

The goodness of the fits is indicated by the small values of the
differences between given and calculated values shown by the line close to
0.0.

It is worth pointing out that the GP map equations representing cyclic
variations are primarily found to be linear. The nonlinearity if any,
arising from the Pad\'{e} type rational terms, has rather small
coefficients as compared to those for the linear terms. It is also noted
that the significant departure of the obtained solutions from a persistent
solution $(X_{t+1}$=$X_{t})$ shows the efficacy of the GP optimization
approach. In the context of specific levels, we observe that the $6^{th}$
level variations show cyclic as well as bursty behavior. Interestingly,
the $7^{th}$ level coefficients show smooth variations; the corresponding
GP equation (Eq. \ref{eqn:Nifty_7thLevel}) is completely linear. On the
contrary the $8^{th}$ level variations show non-smooth variations which
are not bursty like the $6^{th}$ level coefficients.

\begin{figure}
\centering
    {\resizebox{!}{5.5cm}{%
       \includegraphics{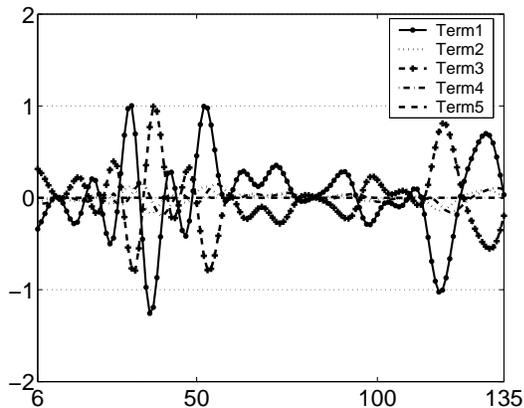}}}
\vspace{-0.1in} \caption{\label{fig:Nifty_L6_Terms}Contributions of
individual terms in the right hand side of Eq. \ref{eqn:Nifty_6thLevel}
for Db4 level-6 wavelet coefficients for Nifty data. It is observed that
the first and third terms are the dominant ones which give rise to the
bursty behavior (Fig. \ref{fig:Nifty_cD6_Fit130}) due to their slight out
of phase dynamics.}
\end{figure}

In order to understand the interplay of different terms giving rise to the
bursty behavior, we have computed the contributions arising from each term
of Eq. \ref{eqn:Nifty_6thLevel} individually. The same is shown in Fig.
\ref{fig:Nifty_L6_Terms}. It is clearly seen that the first and the third
term are the dominant ones, which are slightly out of phase from each
other. The corresponding cancellation is responsible for the bursty
behavior. The dynamical origin of these terms and their modeling is rather
non-trivial, which needs significant investigations.

The map equations will now be used to carry out 1-step out-of-sample
predictions beyond the fitted points. The predictions for level=6, 7 and 8
are shown in Figs. \ref{fig:Nifty_cD6Predict136_1step},
\ref{fig:Nifty_cD7Predict61_1step} and \ref{fig:Nifty_cD8Predict24_1step}
respectively with corresponding NMSE values as 0.04923, 0.03907 and
0.03946 respectively. It can be seen that the 1-step predictions are very
good.

\begin{figure}
\centering
    {\resizebox{!}{5.5cm}{%
       \includegraphics{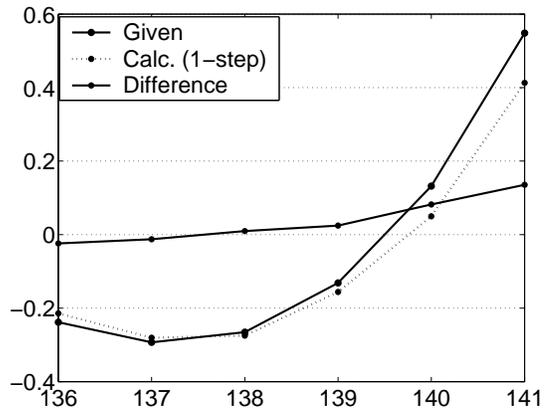}}}
\vspace{-0.1in}
\caption{\label{fig:Nifty_cD6Predict136_1step}Out-of-sample 1-step
predictions using GP solution for Db4 level-6 wavelet coefficients for
Nifty data.}
\end{figure}

\begin{figure}
\centering
    {\resizebox{!}{5.5cm}{%
       \includegraphics{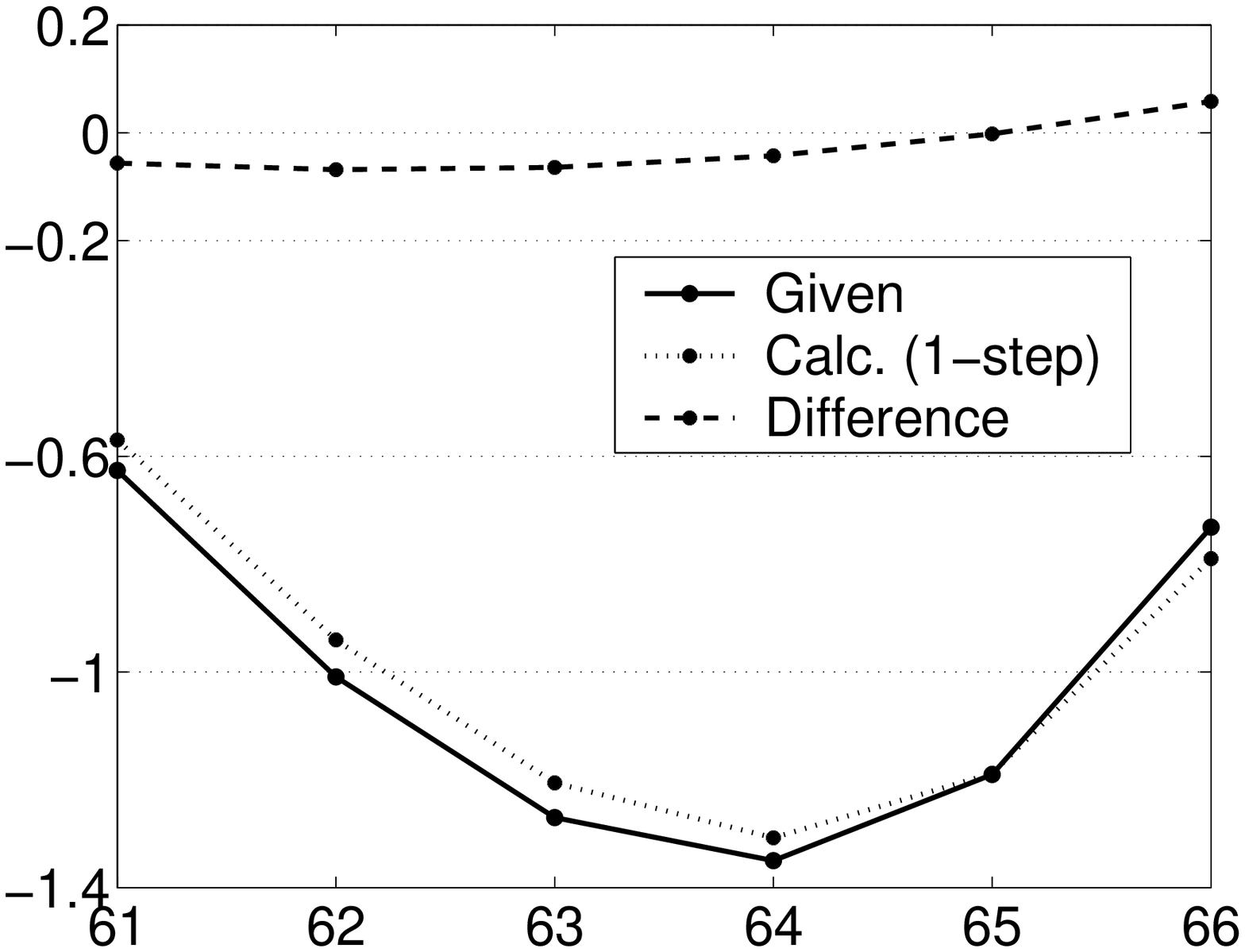}}}
\vspace{-0.1in} \caption{\label{fig:Nifty_cD7Predict61_1step}Out-of-sample
1-step predictions using GP solution for Db4 level-7 wavelet coefficients
for Nifty data.}
\end{figure}

\begin{figure}
\centering
    {\resizebox{!}{5.5cm}{%
       \includegraphics{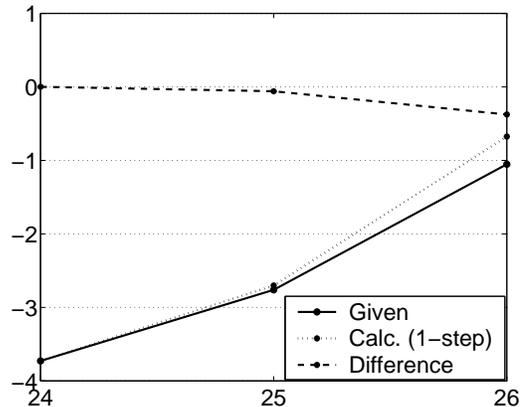}}}
\vspace{-0.1in} \caption{\label{fig:Nifty_cD8Predict24_1step}Out-of-sample
1-step predictions using GP solution for Db4 level-8 wavelet coefficients
for Nifty data.}
\end{figure}

\subsection{Modeling variations in Dow Jones Industrial Average Closing values}
We next consider modeling of Db4 wavelet coefficients for Dow Jones
Industrial Average (DJIA) closing values. Akin to the GP analysis for
Nifty wavelet coefficients, we have found it useful to smoothen the
wavelet coefficients using Cubic Spline for the purpose of GP modeling.
This also makes it easier to compare the cyclic variations in the two time
series considered

The comparison of Db4 forward wavelet coefficients for level=6, 7 and 8
with the Cubic Spline smoothened coefficients are shown in Figs.
\ref{fig:DJIA_cD6_CompareSpline141}, \ref{fig:DJIA_cD7_CompareSpline66}
and \ref{fig:DJIA_cD8_CompareSpline26}.

\begin{figure}
\centering
    {\resizebox{!}{5.5cm}{%
       \includegraphics{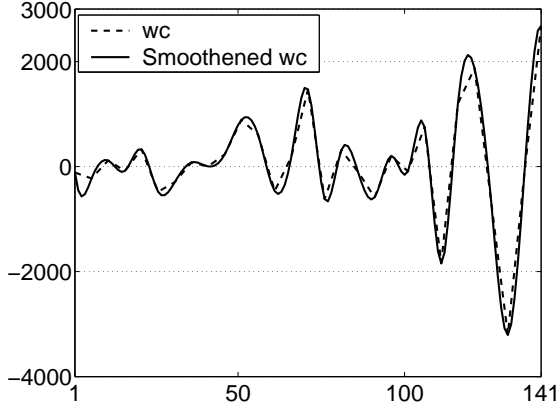}}}
\vspace{-0.1in} \caption{\label{fig:DJIA_cD6_CompareSpline141}Db4 wavelet
coefficients (wc) and Spline interpolated wc for $6^{th}$ level Dow Jones
data.}
\end{figure}

\begin{figure}
\centering
    {\resizebox{!}{5.5cm}{%
       \includegraphics{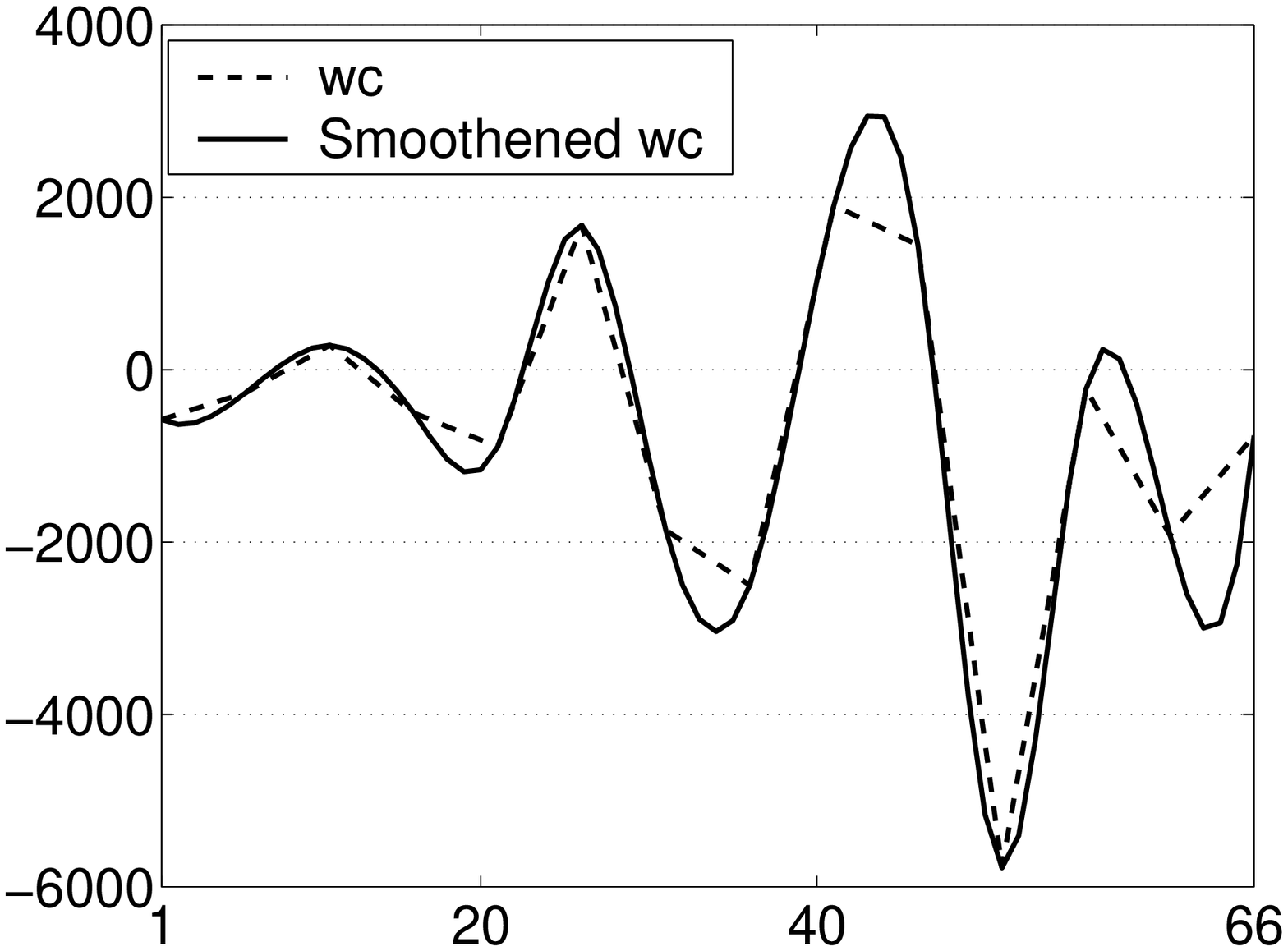}}}
\vspace{-0.1in} \caption{\label{fig:DJIA_cD7_CompareSpline66}Db4 wavelet
coefficients (wc) and Spline interpolated wc for $7^{th}$ level Dow Jones
data.}
\end{figure}

\begin{figure}
\centering
    {\resizebox{!}{5.5cm}{%
       \includegraphics{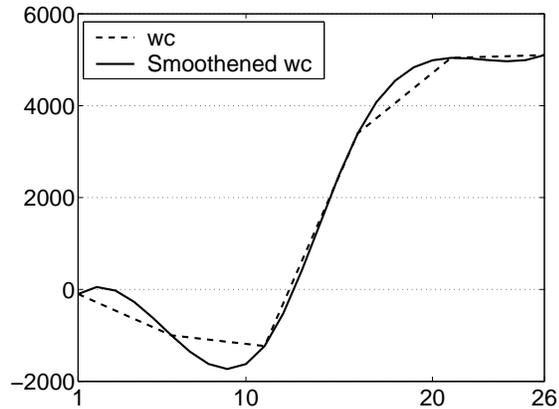}}}
\vspace{-0.1in} \caption{\label{fig:DJIA_cD8_CompareSpline26}Db4 wavelet
coefficients (wc) and Spline interpolated wc for $8^{th}$ level Dow Jones
data.}
\end{figure}

The map equations generated by GP for d=5 and $\tau$=$1$ are having
fitness values 0.998013, 0.99721, and 0.99929 and are shown in Eq.
\ref{eqn:DJIA_6thLevel}, \ref{eqn:DJIA_7thLevel} and
\ref{eqn:DJIA_8thLevel}.

\begin{tiny}
\begin{eqnarray}
X_{t+1}^{(Level=6)}&=&2.4375X_{t}-1.8398X_{t-\tau}+0.12669X_{t-2\tau}+0.2587X_{t-3\tau} \nonumber \\
&+&0.7002+\frac{0.0186(X_{t-4\tau}+248.5)}{X_{t-\tau}} \nonumber \\
&+&\frac{0.6841(1.5X_{t}+125.85)}{X_{t-4\tau}+76.3}
\label{eqn:DJIA_6thLevel}
\end{eqnarray}
\end{tiny}

\begin{tiny}
\begin{eqnarray}
X_{t+1}^{(Level=7)}&=&2.2386X_{t}-1.5X_{t-\tau}+0.0947X_{t-2\tau}+0.1481X_{t-4\tau} \nonumber \\
&-&16.993-\frac{0.71023(X_{t-\tau}+540.519)}{X_{t-2\tau}-X_{t-3\tau}}
\label{eqn:DJIA_7thLevel}
\end{eqnarray}
\end{tiny}

\begin{tiny}
\begin{eqnarray}
X_{t+1}^{(Level=8)}&=&2.20989X_{t}-1.3999X_{t-\tau}+0.16484X_{t-3\tau}+62.9235   \nonumber \\
&-&\frac{0.10989X_{t-\tau}}{X_{t-3\tau}} \label{eqn:DJIA_8thLevel}
\end{eqnarray}
\end{tiny}

As seen in Figs. \ref{fig:DJIA_L6_Fit130}, \ref{fig:DJIA_L7_Fit55} and
\ref{fig:DJIA_L8_Fit18} for level=6, 7 and 8 respectively, the GP fits are
quite good.

\begin{figure}
\centering
    {\resizebox{!}{5.5cm}{%
       \includegraphics{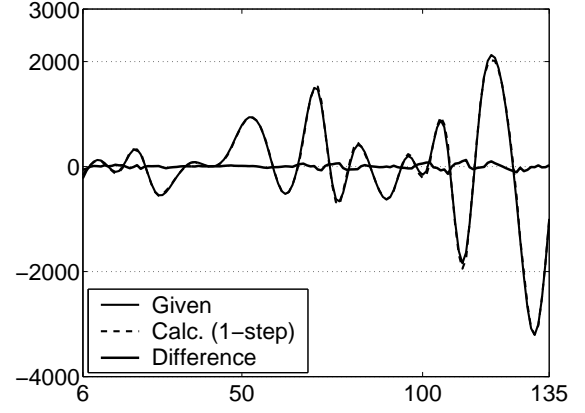}}}
\vspace{-0.1in} \caption{\label{fig:DJIA_L6_Fit130}Fit of the GP solution
for Db4 level-6 Dow Jones wavelet coefficients. The variations have a
bursty character.}
\end{figure}

\begin{figure}
\centering
    {\resizebox{!}{5.5cm}{%
       \includegraphics{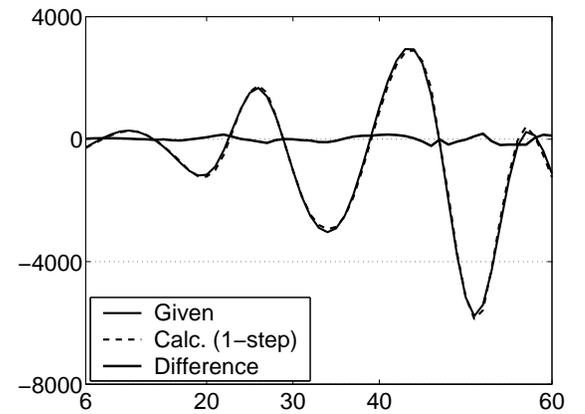}}}
\vspace{-0.1in} \caption{\label{fig:DJIA_L7_Fit55}Fit of the GP solution
for Db4 level-7 Dow Jones wavelet coefficients. The amplitude of the
cyclic variations is seen to increase with time.}
\end{figure}

\begin{figure}
\centering
    {\resizebox{!}{5.5cm}{%
       \includegraphics{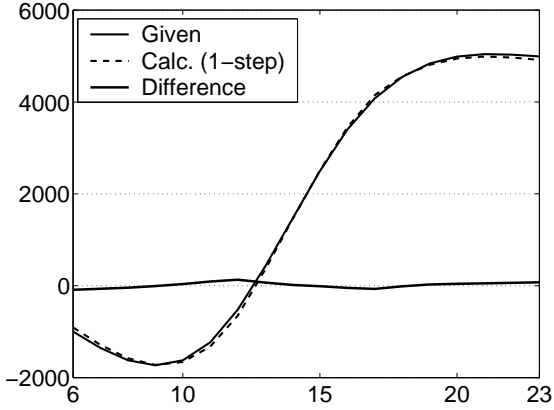}}}
\vspace{-0.1in} \caption{\label{fig:DJIA_L8_Fit18}Fit of the GP solution
for Db4 level-8 Dow Jones wavelet coefficients. The variations show a step
like behavior.}
\end{figure}

Similar to the Nifty analysis, the GP equations are primarily of linear
type having non-linear terms of Pad\'{e} type. Eq. \ref{eqn:DJIA_6thLevel}
is primarily linear. The effect of nonlinearity as seen from the Pad\'{e}
terms is different from the Nifty behavior. Eq. \ref{eqn:DJIA_7thLevel}
shows a very interesting behavior. If difference between two consecutive
data points are small, then the Pad\'{e} term gives a strong contribution,
which decreases as the differences increase. Eq. \ref{eqn:DJIA_8thLevel}
representing level 8 coefficients is again mostly linear.

We then use these map equations and carry out 1-step out-of-sample
predictions beyond the fitted points. These predictions are found to be
very good as is reflected from their small NMSE values, 0.001787
(level=6), 0.002379 (level=7) and 0.0004981 (level=8). The predictions are
shown for level=6, 7 and 8 in Figs. \ref{fig:DJIA_L6_GPPredict130_1step},
\ref{fig:DJIA_L7_GPPredict55_1step} and
\ref{fig:DJIA_L8_GPPredict18_1step} respectively and are found to be
excellent.

\begin{figure}
\centering
    {\resizebox{!}{5.5cm}{%
       \includegraphics{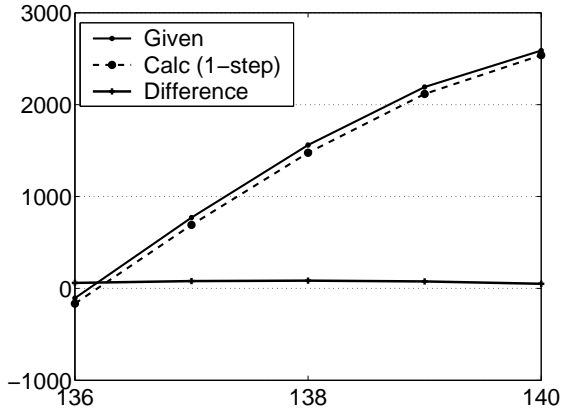}}}
\vspace{-0.1in}
\caption{\label{fig:DJIA_L6_GPPredict130_1step}Out-of-sample 1-step
predictions using GP solution for Db4 level-6 Dow Jones wavelet
coefficients.}
\end{figure}

\begin{figure}
\centering
    {\resizebox{!}{5.5cm}{%
       \includegraphics{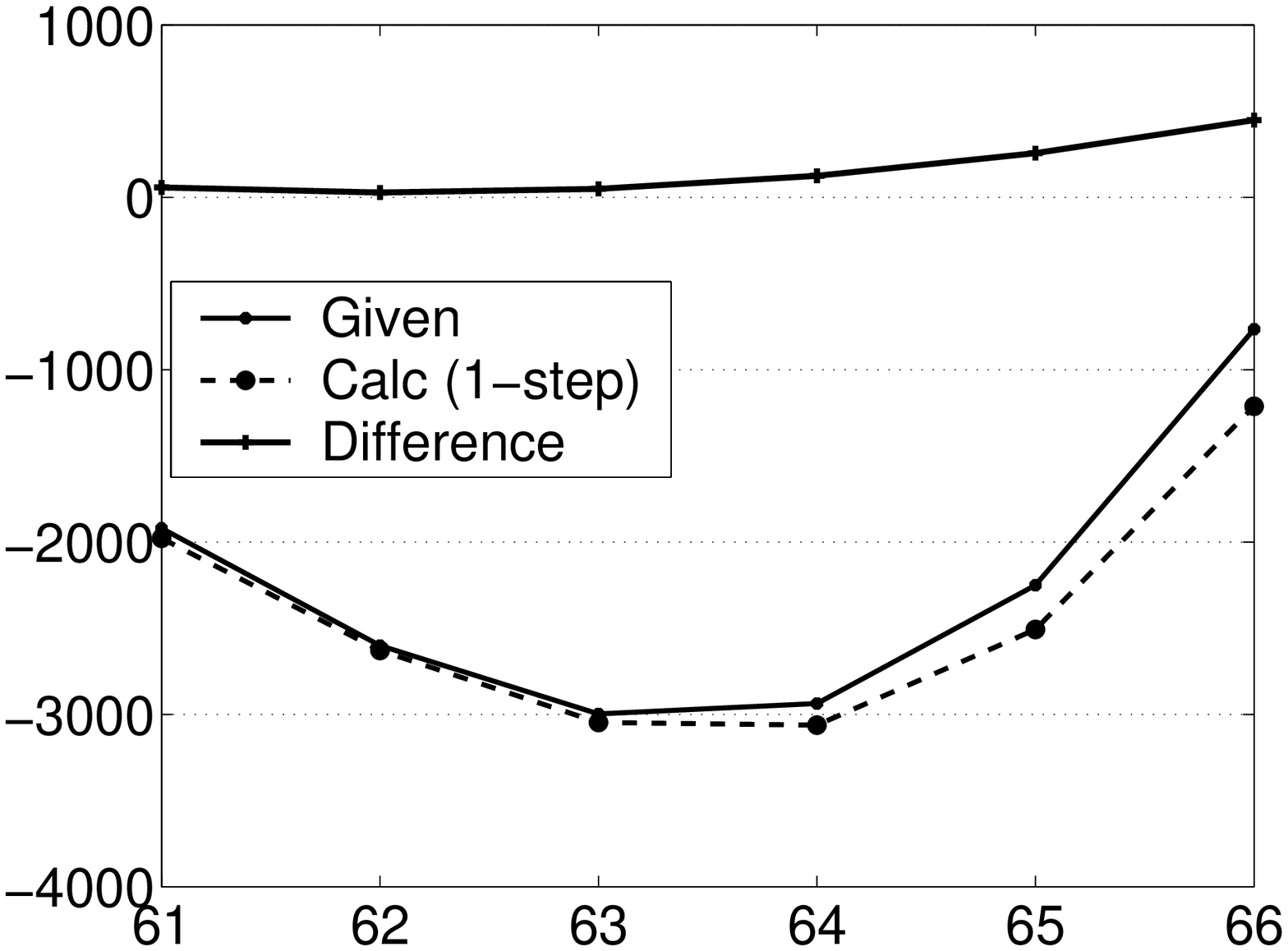}}}
\vspace{-0.1in}
\caption{\label{fig:DJIA_L7_GPPredict55_1step}Out-of-sample 1-step
predictions using GP solution for Db4 level-7 Dow Jones wavelet
coefficients.}
\end{figure}

\begin{figure}
\centering
    {\resizebox{!}{5.5cm}{%
       \includegraphics{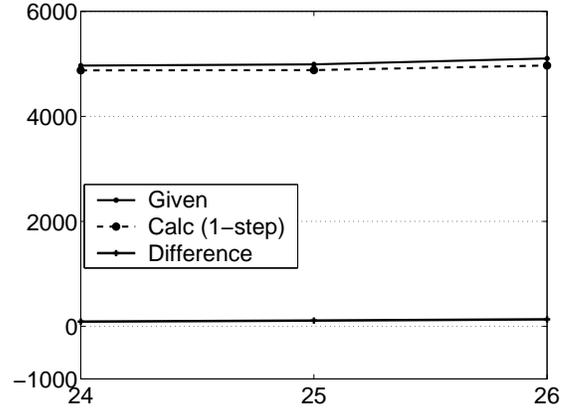}}}
\vspace{-0.1in}
\caption{\label{fig:DJIA_L8_GPPredict18_1step}Out-of-sample 1-step
predictions using GP solution for Db4 level-8 Dow Jones wavelet
coefficients.}
\end{figure}

\section{Conclusion}
In conclusion, we have illustrated a wavelet based approach to separate
stochastic and structured variations in non-stationary time series.
Modeling different aspects e.g., fluctuations and trend of these time
series is a challenging task. It becomes particularly difficult when the
fluctuations comprise of random, cyclic and transient variations at
multiple scales. The fact that wavelet transform possesses
multi-resolution analysis capability, has opened the way to isolate
variations at different scales. We have taken advantage of this ability of
wavelets to study and model cyclic variations of the financial time
series, which are known to be non-stationary. Genetic programming models
the cyclic behavior well through crisp dynamical equations. One step
predictions have been carried through and these are found to be quite
accurate.

Apart from studying other physical time series, it will be nice to combine
the present approach with random matrix based ones for the purpose of
pinpointing emergence of cyclic behavior. As has been mentioned earlier,
random matrix approach has indicated correlation between group of
companies in financial time series, which can lead to cyclic or structured
variations apparent in the present analysis. Hence, it will be of deep
interest to see if these two can be interrelated.

Amit Verma is thankful to Physical Research Laboratory for providing him a
project traineeship during which part of this work was done.


\end{document}